\documentclass[prb,
superscriptaddress,showpacs,amsmath,amssymb]{revtex4}
\usepackage{amsfonts}
\usepackage{bm}
\usepackage{verbatim}

\usepackage{graphicx}

\begin{document}
\title{Gravity through the prism of condensed matter physics}

\author{G.E.~Volovik}
\affiliation{Low Temperature Laboratory, Aalto University,  P.O. Box 15100, FI-00076 Aalto, Finland}
\affiliation{Landau Institute for Theoretical Physics, acad. Semyonov av., 1a, 142432,
Chernogolovka, Russia}

\date{\today}

\begin{abstract}
{In the paper "Life, the Universe, and everything—42 fundamental questions",
Roland Allen and Suzy Lidstr\"om presented personal selection of the fundamental questions. Here, based on the condensed matter experience, we suggest the answers to some questions concerning the vacuum energy, black hole entropy and the origin of gravity. In condensed matter we know both the many-body phenomena emerging on the macroscopic level and the microscopic (atomic) physics, which generates this emergence. It appears that the same macroscopic phenomenon may be generated by essentially different microscopic backgrounds. This points to various possible directions in study of the deep quantum vacuum of our Universe.}
\end{abstract}

\maketitle

\tableofcontents

\section{Introduction: 42 fundamental questions}

Below are the questions presented in the paper by Allen and
Lidstr\"om “Life, the Universe, and everything--42 fundamental questions”.\cite{Perspective}

1. {\bf Why does conventional physics predict a cosmological constant that is vastly
too large?}

2. {\bf What is the dark energy?}

3. {\bf How can Einstein gravity be reconciled with quantum mechanics?}

4. {\bf What is the origin of the entropy and temperature of black holes?}

5. {\bf Is information lost in a black hole?}

6. {\bf Did the Universe pass through a period of inflation, and if so how and why?}

7. {\bf Why does matter still exist?}

8. {\bf What is the dark matter?}

9. {\bf Why are the particles of ordinary matter copied twice at higher energy?}

10. {\bf What is the origin of particle masses, and what kind of masses do
neutrinos have?}

11. {\bf Does supersymmetry exist, and why are the energies of observed particles so
small compared to the most fundamental (Planck) energy scale?}

12. {\bf What is the fundamental grand unified theory of forces, and why?}

13. {\bf Are Einstein relativity and standard field theory always valid?}

14. Is our Universe stable?

15. {\bf Are quarks always confined inside the particles that they compose?}

16. What are the complete phase diagrams for systems with nontrivial forces,
such as the strong nuclear force?

17. {\bf What new particles remain to be discovered?}

18. {\bf What new astrophysical objects are awaiting discovery?}

19. What new forms of superconductivity and superfluidity remain to be
discovered?

20. What new topological phases remain to be discovered?

21. What further properties remain to be discovered in highly correlated
electronic materials?

22. What other new phases and forms of matter remain to be discovered?

23. What is the future of quantum computing, quantum information, and other
applications of entanglement?

24. What is the future of quantum optics and photonics?

25. {\bf Are there higher dimensions, and if there is an internal space, what is its
geometry?}

26. {\bf Is there a multiverse?}

27. {\bf Are there exotic features in the geometry of spacetime, perhaps including
those which could permit time travel?}

28. How did the Universe originate, and what is its fate?

29. {\bf What is the origin of spacetime, why is spacetime four-dimensional, and why
is time different from space?}

30. {\bf What explains relativity and Einstein gravity?}

31. {\bf Why do all forces have the form of gauge theories?}

32. {\bf Why is Nature described by quantum fields?}

33. Is physics mathematically consistent?

34. {\bf What is the connection between the formalism of physics and the reality of
human experience?}

35. What are the ultimate limits to theoretical, computational, experimental, and
observational techniques?

36. What are the ultimate limits of chemistry, applied physics, and technology?

37. What is life?

38. How did life on Earth begin and how did complex life originate?

39. How abundant is life in the Universe, and what is the destiny of life?

40. How does life solve problems of seemingly impossible complexity?

41. Can we understand and cure the diseases that afflict life?

42. What is consciousness?

Condensed matter has its answers to the most of the questions, right or wrong. In some cases there are several answers to the same question, which may contradict to each other. For example, there are at least 6 different answers to the question  \#30 concerning the origin of  relativity and Einstein gravity discussed in Sec. \ref{gravity}.  However, they may open different directions in our research. 

 The questions which are (at least partially) touched in these notes are given in boldface, they are mostly related to gravity. The answers to some other questions can be found in Ref. \cite{Volovik2020}.

\section{Cosmological constant and vacuum energy}
\label{VacuumEnergy}

According to standard physics, the vacuum has an enormous energy density $\rho_{\rm vac}$. A positive contribution comes from the
zero-point energy of bosonic fields, such as electromagnetic field, and a negative contribution
comes from the fermionic fields -- from the so called Dirac vacuum, and there is no reason why they should cancel.\cite{Weinberg1989}
Again according to standard physics, $\rho_{\rm vac}$ should act as a gravitational source
-- effectively an enormous cosmological constant $\Lambda_{\rm vac}$. With the Planck energy scale $E_{\rm Planck}$ providing a natural cutoff, $\Lambda_{\rm vac}\sim E_{\rm Planck}^4$ is roughly 120 orders of magnitude
larger than is compatible with observations.
According to Bjorken, this is the oft-repeated mantra that "no one has any idea as to why the cosmological constant is so small".\cite{Bjorken2011}

\subsection{Nullification of the energy of the ground state in condensed matter}
\label{NullificationCondMat}

However, anyone who is familiar with condensed matter physics can immediately find a loophole in this logic, simply by considering the ground state of the liquid or solid quantum material consisting for example of the atoms of one sort. This non-relativistic many body system is described by the thermodynamic potential $\epsilon -\mu n$, where $\epsilon$ is the energy density, $n$ is particle density and $\mu=d\epsilon/dn$ is the chemical potential. According to the Gibbs-Duhem identity,  at zero temperature one has 
 \begin{equation}
\epsilon -\mu n=-P\,,
\label{EoS}
\end{equation}
where $P$ is pressure.  This  thermodynamic potential $\epsilon -\mu n$ is  equivalent to the vacuum energy $\rho_{\rm vac}$, because it obeys the same equation of states:
 \begin{equation}
 \rho_{\rm vac}=-P_{\rm vac}\,.
\label{EquationOfState}
\end{equation}

The low energy modes of condensed matter,  such as phonons and electronic excitations,  are described in terms of bosonic and  fermionic quantum fields.\cite{AGD} These fields have zero-point energies, which contribute to the thermodynamic potential and do not cancel each other. Nevertheless,  in the absence of environment, i.e. at zero pressure, the thermodynamic potential of the system is exactly zero,   $\rho_{\rm vac}\equiv \epsilon -\mu n=-P=0$. 

What happens with the diverging contribution of the zero-point-energies?  It is automatically cancelled by the microscopic (atomic) degrees of freedom due to  the Gibbs--Duhem identity. According to Einstein, "thermodynamics ... is the only physical theory of universal content", and thus
the thermodynamic identity should be equally applicable to equilibrium ground state of any condensed matter system and to the physical vacuum, whatever is its content. The thermodynamics dictates that the enormous contribution of zero-point energies of quantum fields to the vacuum energy should be automatically cancelled by the microscopic (now trans-Planckian) degrees of freedom of the quantum vacuum. If the vacuum is in full equilibrium, one has $\rho_{\rm vac}=-P_{\rm vac}=0$, and this does not depend on the physical content of the quantum vacuum.

\subsection{Nullification of the vacuum energy in equilibrium}
\label{NullificationVacEnergy}

Such condensed matter scenario of nullification of the vacuum energy in the full equilibrium is reproduced in the so-called $q$-theory.\cite{KlinkhamerVolovik2008a,KlinkhamerVolovik2008b} Here the vacuum variable $q$, which corresponds to the density $n$, is described by the 4-form field introduced by Hawking for study the vacuum energy,\cite{Hawking1984} $F_{\alpha\beta\mu\nu}=q e_{\alpha\beta\mu\nu}\sqrt{-g}$. The equation of state of the vacuum is given by the same equations (\ref{EoS}) and (\ref{EquationOfState}):
 \begin{equation}
\epsilon_{\rm vac}(q)  -\mu q \equiv  \rho_{\rm vac}=-P_{\rm vac} \,\,,\,\, \mu= \frac{d\epsilon_{\rm vac}}{dq}\,.
\label{EoSq}
\end{equation}
It is important that the role of the vacuum energy density, which enters the Einstein equation as cosmological constant $\Lambda_{\rm vac}$, is played not by the energy density $\epsilon_{\rm vac}(q)$, but by the analog of the thermodynamic potential
$\Lambda_{\rm vac}=\epsilon_{\rm vac}(q)- \mu q$. This modification of the vacuum energy density was missing in the Hawking paper\cite{Hawking1984}, and this  was later corrected in the followed papers.\cite{Duff1989,Wu2008}

In the full equilibrium, the huge energy density $\epsilon_{\rm vac}(q)$ is fully compensated by the "counterterm" $-\mu q$. This compensation takes place due to the conventional laws of thermodynamics. The condensed matter systems at zero temperature (such as superfluid $^4$He and superfluid phases of $^3$He) demonstrate how this nullification works without fine tuning, see Eq.(\ref{EoS}).
In the absence of external pressure, the large (atomic) value of the energy density $\epsilon$ is fully compensated  by the "counterterm" $-\mu n$.

\subsection{Relaxation of the vacuum energy}
\label{RelaxationVacEnergy}

In the expanding Universe,  the vacuum is out of equilibrium and/or is contaminated by matter. As a result, the vacuum energy deviates from its equilibrium value in the same manner as it happens in condensed matter at finite temperature. In the early Universe the vacuum energy density could be huge, even of Planck scale, $\rho_{\rm vac}\sim E_{\rm Planck}^4$ (here $E_{\rm Planck}^2=1/G$, where $G$ is Newton constant, and we use units $\hbar=c=1$). But in the present epoch, the Universe is old and is close to equilibrium. Its expansion is slow,  $H\ll E_{\rm Planck}$, where $H$ is the Hubble parameter, and the matter is highly dilute. That is why the vacuum energy density (the dark energy) only slightly deviates from its zero value, being on the order of the perturbations of the vacuum state caused by expansion and matter. Now it is on the order of the density of the dilute matter 
$\rho_{\rm vac}\sim \rho_{\rm mat}$. In principle, the condensed matter physics may suggest that the dark energy is just the response of the quantum vacuum to the gravitating matter.

The condensed matter also demonstrates, how the initial states with large $\rho_{\rm vac}$ relax to the equilibrium state with $\rho_{\rm vac}=0$. In superconductors, such relaxation is typically accompanied by rapid oscillations.\cite{VolkovKogan1974,Barankov2004,Yuzbashyan2005,Yuzbashyan2008,Gurarie2009}
Similar oscillations after inflation were obtained by Starobinsky,\cite{Starobinsky1980} and they were also obtained in $q$-theory.\cite{KlinkhamerVolovik2008b} 
These spacetime-dependent perturbations behave gravitationally as a pressureless perfect fluid, which may serve as a candidate for dark matter.\cite{KlinkhamerVolovik2017}

The $q$-theory gives the following decay of the vacuum energy density after averaging over the fast oscillations:  
$\bar\rho_{\rm vac}\sim E_{\rm Planck}^2/t^2$. It changes from $\bar \rho_{\rm vac}\sim E_{\rm Planck}^4$ at the Planck time $t=t_{\rm Planck}=1/E_{\rm Planck}$ to about the present value at present time. The time dependent Newton "constant" $G(t)$ experiences the oscillating decay approaching its value $G(t=\infty)$ in the equilibrium quantum vacuum:
\begin{equation}
\frac{1}{G(t)} - \frac{1}{G(\infty)} \sim \frac{E_{\rm Planck}}{t} \, \sin(E_{\rm Planck} t)\,.
 \label{G}
\end{equation}
.

.

\subsection{Coexistence of phases and Multiverse}
\label{Coexistence}

In principle, the equilibrium Universe can live on the coexistence curve of two or more phases of
the quantum vacuum, which is typical for condensed matter systems. In particle physics such multiverse (question \#26) corresponds to the Multiple Point Principle.\cite{Nielsen1994,Nielsen2016,Kawai2022} In condensed matter, the coexistence of different quantum vacua (ground states) can be regulated by the exchange of the global
charges between the vacua, such as spin and particle number. In particle physics it can be regulated by exchange of baryonic and leptonic charges. As a result, each coexisting vacuum will have the nonzero baryonic charge, which in turn induces the baryonic asymmetry of matter.\cite{Volovik2004} 
This is the condensed matter answer to the question  \#7.

\subsection{Conclusion}
\label{NullificationConclusion}

Finally the answer to the question  \#1 "Why does conventional physics predict a cosmological constant that is vastly too large?" is the following. The so-called "conventional physics" ignores the condensed matter lesson that in the full equilibrium the diverging zero-point energy of quantum fields is cancelled by  the atomic (trans-Planckian) degrees of freedom. The reason for cancellation is the general laws of thermodynamics, which are universal and thus the same for relativistic and non-relativistic vacua (ground states). In general, the cosmological "constant" (the energy density of the vacuum) can be huge, but in the equilibrium vacuum state it is zero without fine tuning.

According to condensed matter, this is also valid for the metastable vacuum state -- the false vacuum. In condensed matter both the true vacuum and the false vacuum have zero energy density, $\epsilon - \mu n=0$, although they have different values of the chemical potential $\mu$. This should be taken into account in theories, in which the cosmic inflation is an effect of a false vacuum decaying into the true vacuum.  
 
\section{Entropy of horizon}
\label{Horizon}

\subsection{De Sitter state as perturbed vacuum, and its local temperature}
\label{deSitterPerturbed}

The de Sitter state of expanding Universe is an example of the perturbed quantum vacuum. It has nonzero energy density $\rho_{\rm dS}=\frac{3H^2}{8\pi G}$, where $H$ is the Hubble parameter of expansion.

Different approaches suggest that this excited state of the original Minkowski vacuum can be considered as the thermal state of the quantum vacuum.
However, the value of the temperature is different in different apporaches.
In the traditional approach this temperature is determined by the Hawking radiation from the cosmological  event horizon. 

The rate of the Hawking radiation from the cosmological horizon or from the black hole horizon can be calculated using the mechanism of quantum mechanical tunneling. According to Carlip:\cite{Carlip2014} "Perhaps the most intuitive is to say that quantum mechanics allows particles to tunnel out through the event horizon".
This tunneling process of Hawking radiation was considered for the condensed matter analog of black hole in Ref.\cite{Volovik1999}, and for the astrophysical black holes in Refs.\cite{Srinivasan1999,Wilczek2000}. 
In both cases the rate of the tunneling process is characterized by the Hawking temperature, which is determined by the (effective) gravity at the horizon.

The same result was obtained for the rate of quantum tunneling from the de Sitter horizon,\cite{Volovik2009a} where the corresponding temperature (the Gibbons-Hawking temperature) is $T_{\rm GH}=H/2\pi$. However, by the same tunneling method it was found that locally the observer measures the temperature, which is twice larger, $T=H/\pi$.
This temperature, for example, determines the ionization rate of atom in the de Sitter environment,\cite{Volovik2009a,Maxfield2022} which is the local process without any relation to the cosmological horizon. The same temperature determines the rate of the other local processes,  such as the decay rate of the composite particle, which is stable in the flat spacetime.\cite{Volovik2009b}  
Also the probability of evaporation of $^4$He atoms from superfluid liquid is $\propto \exp(-\pi |\mu|/H)$, where $\mu$ is the chemical potential. For condensed matter physicist this is the strong evidence that $T=H/\pi$ is the temperature  of the de Sitter thermal bath.

The calculations of the decay rate of the local processes are within the well-known physics, and do not require the existence of the horizon and the ultraviolet physics. The existence of the local temperature suggests  that it is the real temperature of the de Sitter thermal state, $T_{\rm dS}=H/\pi$. Then what to do with the temperature $T_{\rm GH}=H/2\pi$, which determines the Hawking radiation from the cosmological horizon? The condensed matter gives the answer to this question too. 

From the condensed matter point of view, the Hawking radiation is the process of coherent creation of the entangled pair of particles, inside and outside the cosmological horizon. The rate of creation can be written in two different ways depending on the quality of the observer:
\begin{equation}
w\sim \exp{ \left(-\frac{2E}{T_{\rm dS}}\right)}=\exp{ \left(-\frac{E}{T_{\rm GH}}\right)}\,.
 \label{DoubleHawking}
\end{equation}
The sub-luminal observer can detect only single particle -- the particle which appears inside the cosmological horizon. For this observer the process is described by the right-hand side of Eq.(\ref{DoubleHawking}) with temperature  $T_{\rm GH}=H/2\pi$.  In the condensed matter experiments with the analogs of the cosmological horizon, the experimentalist is "superluminal" (supersonic) and can detect the simultaneous creation of both particles.  For such external observer the process of Hawking radiation is described by the temperature $T_{\rm dS}=H/\pi$ on  the left-hand side of Eq.(\ref{DoubleHawking}).  Similar (but different) mechanism of doubling the Hawking temperature due to entanglement was suggested by ’t Hooft for black hole.\cite{tHooft2022} 

In the local ionization process only single particle is created, and thus the rate of ionization measured both by the sub-luminal and the super-luminal observers is characterized by the same local temperature, $w\propto \exp{ \left(-\frac{E}{T_{\rm dS}}\right)}$.  Since in this case there is no disagreement between the local and global observers,  it is natural to assume that it is the local temperature $T_{\rm dS}=H/\pi$, which determines the thermodynamics of the de Sitter thermal state.

\subsection{Entropy of expanding Universe}
\label{deSitterEntropy}

Due to the symmetry of the de Sitter spactime, the local temperature $T=T_{\rm dS}$ is the same for all static observers. The energy density at any point of the de Sitter space, when expressed in terms of the local temperature, can be considered as the density of the thermal energy in de Sitter space:
\begin{equation}
 \rho_{\rm dS}(T)=\frac{3}{8\pi G}H^2=\frac{3\pi}{8G}T^2\,.
\label{dSEnergyDensity}
\end{equation}
From this thermal energy one can obtain the local entropy density in the de Sitter thermodynamics:
\begin{equation}
s_{\rm dS}(T)= - \frac{\partial F}{\partial T} =\frac{3\pi}{4G}T=\frac{3}{4G}H\,,
\label{dSEntropyDensity}
\end{equation}
where the free energy density $F(T)=\rho_{\rm dS}(T) - T d\rho_{\rm dS}/dT=- \rho_{\rm dS}(T)$. 

The total entropy in the volume surrounded by the cosmological horizon with radius $R=1/H$ is
\begin{equation}
S_{\rm dS}=\frac{4\pi R^3}{3} s _{\rm dS}= \frac{\pi}{GH^2}=\frac{A}{4G} \,,
\label{dSEntropy}
\end{equation}
where $A$ is the horizon area.
This corresponds to the Gibbons-Hawking entropy of the cosmological horizon. But in this approach,  the thermodynamic entropy comes from the local entropy of the de Sitter thermal state, rather than from the cosmological horizon. This is related to the question  \#4 concerning the black hole entropy, which we discuss in Sec.\ref{BH}. 

\subsection{Two-fluid dynamics of vacuum and matter}

Till now we considered the pure de Sitter state, and did not take into account the matter degrees of freedom. From the condensed matter point of view the expanding Universe has some similarity with 
superfluid liquid: the combined dynamics of vacuum and matter has signatures of the Landau-Tisza two-fluid model. The role of the superfluid component of the liquid is played by the gravitational degrees of freedom, while matter corresponds to the normal component of the liquid (system of quasiparticles).

The vacuum may have its own temperature and entropy.\cite{Padmanabhan2020} 
In the expanding Universe, the vacuum degrees of freedom typically have much higher entropy density than the matter degrees of freedom.\cite{Volovik2023}  The quasi-equilibrium states of the expanding Universe may have different temperatures for gravitational vacuum and for matter degrees of freedom.\cite{Vergeles2023}

\subsection{Gibbs-Duhem relation in de Sitter thermodynamics}

In de Sitter state, the "superfluid" component contains the pair of the  thermodynamically conjugate variables:\cite{KlinkhamerVolovik2008c}  the gravitational coupling $K=\frac{1}{16\pi G}$ and the scalar Riemann curvature  ${\cal R}$,
 which enter the Einstein-Hilbert action
\begin{equation}
S_{\rm{GR}}  = \int d^4 x \sqrt{-g} \,K {\cal R} \,.
\label{GR}
\end{equation}
The de Sitter thermodynamics obeys the corresponding analog of the Gibbs-Duhem relation:
 \begin{equation}
T_{\rm dS}s_{\rm dS}=\rho_{\rm dS}+P_{\rm dS} -K{\cal R}=-K{\cal R}\,.
\label{GibbsDuhem}
\end{equation}
This equation is satisfied, because the  "superfluid" component obeys the vacuum equation of states, $P_{\rm dS}=w\rho_{\rm dS}$ with $w=-1$; and the scalar curvature is ${\cal R}=-12H^2=-12\pi^2 T^2$.

Eq.(\ref{GibbsDuhem}) suggests that one may introduce the effective pressure, which is modified by gravitational degrees of freedom, $\tilde P_{\rm dS}= P_{\rm dS} -K{\cal R}$. Then one obtains the conventional Gibbs-Duhem relation:
\begin{equation}
T_{\rm dS}s_{\rm dS}=\rho_{\rm dS}+\tilde P_{\rm dS}\,.
\label{GibbsDuhem2}
\end{equation}
The modified pressure $\tilde P_{\rm dS}$ of the de Sitter thermal state obeys the equation of state of stiff matter in the cosmological model introduced by Zel'dovich,\cite{Zeldovich1962} $\tilde P_{\rm dS}=w\rho_{\rm dS}$ with $w=1$. For stiff matter, the speed of sound $s$ is equal to the speed of light, $s^2=c^2 dP/d\epsilon=c^2$.  

Also, due to the gravitational degrees of freedom, the de Sitter state has common properties with the non-relativistic Landau Fermi liquid, where the thermal energy is proportional to $T^2$,
and the entropy density is linear in $T$.

\subsection{Entropy of black hole}
\label{BH}

 The same thermodynamic approach can be applied to black holes.
As distinct from the de Sitter state, the black hole is the compact object, and its thermodynamics connects the global parameters, such as mass $M$, entropy of horizon $S_{\rm BH}$, total electric charge $Q$ and total angular momentum $J$. This global thermodynamics can be described by the integral form of the  Gibbs-Duhem relation. 

For the Schwarzschild black hole the local Gibbs-Duhem relation in Eq.(\ref{GibbsDuhem}) becomes: 
\begin{equation}
T_{\rm BH} s({\bf r}) =\epsilon({\bf r})- K{\cal R}({\bf r})\,,
\label{LocalGibbs}
\end{equation}
where $T_{\rm BH}=\frac{1}{8\pi MG}$ is the Hawking temperature.
Using  the  scalar Riemann curvature coming from the central singularity,\cite{Balasin1993}  
${\cal R}({\bf r})=8\pi MG\,\delta({\bf r})$, and mass density $\epsilon({\bf r})=M\,\delta({\bf r})$, one obtains the entropy density and the total entropy of the Schwarzschild black hole:
\begin{equation}
 s({\bf r}) = \frac{M}{2T_{\rm BH}}\delta({\bf r})\,\,, \,\, S_{\rm BH}=\int d^3 r\,  s({\bf r}) = \frac{A}{4G} \,,
\label{BlackGibbs}
\end{equation}
where $A$ is the area of the event horizon.

This demonstrates that the Bekenstein-Hawking entropy of black hole also comes from the local thermodynamics. The entropy is concentrated in the central singularity, which has the mass component ("normal component") and the gravity component ("superfluid component").  The similar result was obtained in Ref. \cite{Lam2023}.  The concentration of the entropy in the central singularity  suggests the large information stored in this singularity. This is the condensed matter answer to the question  \#5.

\subsection{Black hole singularity problem}
\label{Singularity}

In condensed matter physics the problems of singularities are typically well resolved. We have the hierarchy of the length scales, and singularity formed at one scale is resolved within the shorter length scale. Example is the structure of the vortex in superfluids. At large scale, the vortex in $^3$He-B has $\delta$-function singularity in vorticity, $\nabla \times {\bf v}_{\rm s}=(\hbar/2m) \, \delta_2({\bf r})$.  This singularity is resolved at the scales of the coherence length. 

Interestingly, on this short scale the axial symmetry of the vortex can be spontaneously broken. In superfluid $^3$He-B, the vortex splits into two half-quantum vortices separated by the domain wall.\cite{Thuneberg1986,VolovikSalomaa1985,Kondo1991,Silaev2015}
This is the cosmic analog of the Alice strings separated by the Kibble-Lazarides-Shafi walls.\cite{Kibble1982,Makinen2019,Makinen2023} It is not excluded, that the similar symmetry breaking may happen within the central singularity inside the black hole. In the sub-Planckian length scale the central singularity may spontaneously loose the spherical symmetry, forming for example a kind of  rigid top. This in turn may influence the shape of the event horizon.

\section{Scenarios of emergent gravity}
\label{gravity}

What is the origin of gravity from the point of view of condensed matter? It happens that there are at least 6 different scenarios of emergent gravity, which are supported by the condensed matter examples, and it is not clear which of them (if any) is preferred by Nature. 
 
 \subsection{Tetrads as bilinear forms of fermion operators}
 \label{gravity1}
 
This is the Akama-Diakonov theory. It is written in terms of the fundamental spinor matter,\cite{Akama1978,Wetterich2004} and also with the spin connection included as an independent gauge
variable,\cite{Diakonov2011} see also review in Ref.\cite{Sindoni2012}. The gravitational tetrads $ e^a_\mu$ emerge as the order parameter of symmetry breaking phase transition, the vacuum expectation value of the bilinear form of fermionic operators:
 \begin{equation}
 e^a_\mu=<\hat E^a_\mu>\,\,,\,\, 
 \hat E^a_\mu = \frac{1}{2}\left( \Psi^\dagger \gamma^a\partial_\mu  \Psi -  \Psi^\dagger\overleftarrow{\partial_\mu}  \gamma^a\Psi\right) \,.
\label{TetradsFermions}
\end{equation}
 The metric field is the bilinear combination of the tetrad fields,
$g_{\mu\nu}=\eta_{ab}e^a_\mu e^b_\nu$, 
and thus in this quantum gravity the metric is the fermionic quartet.
The symmetry breaking scheme here is $L_{L}\times L_{S}\rightarrow L_{L+S}$, where $L_{L}$ and $L_{S}$ are two separate  symmetries under Lorentz rotations of the coordinate and spin space respectively (the spin connection is the gauge field of the $L_{S}$ group). These two symmetries are broken to the diagonal subgroup -- the Lorentz group of the combined  rotations in two spaces,  $L_{L+S}$. 

This scenario is supported by the condensed matter example,  in which vielbein emerges as the order parameter and metric emerges as fermionic quartet. This takes place for the spin-triplet $p$-wave superfluid $^3$He-B, where the effective gravity emerges as a result of the symmetry breaking $SO(3)_{L}\times SO(3)_{S}\rightarrow SO(3)_{L+S}$.\cite{Volovik1990} Here instead of the orbital and spin Lorentz groups one has three-dimensional rotations in the orbital and spin spaces. The symmetry breaking is described using the conventional BCS four-fermion theory, which is similar to Nambu-Jona-Lasinio type in Ref.\cite{Terazawa1977}.

 \subsection{Tetrads from Weyl points}
 \label{gravity2}
 
In this scenario, tetrads emerge together with Weyl fermions and gauge fields in the vicinity of the topological Weyl points in the energy spectrum of fermionic quasiparticles.\cite{Volovik2003,Horava2005}  For example, the expansion of the $2\times 2$ Hamiltonian for fermionic quasiparticles near the Weyl point at ${\bf p}={\bf p}^0$ gives 
\begin{equation}
 H \approx e^i_a \sigma^a(p_i -p_i^0) \,.
\label{TetradsWeyl}
\end{equation}
Here $\sigma^a$ are Pauli matrices; the spacetime dependent matrix $e^i_a $ plays the role of the effective triad field; and the spacetime dependent vector  ${\bf p}^0$ plays the role of the vector potential of the effective $U(1)$ gauge field. This scenario takes place in the Weyl superfluid $^3$He-A and now is intensively discussed in Weyl semimetals. 

In this scenario, it is possible to simulate the black hole\cite{Volovik2016} and white hole\cite{Wilczek2020} horizons and closed timelike curves.\cite{NissinenVolovik2017}
The gapless fermions allows us to simulate the zero-charge phenomenon in quantum electrodynamics\cite{RantanenEltsov2023,Volovik1988} (the logarithmic behavior of the fine structure constant, $1/\alpha \propto \ln(\Delta/T)$, where the gap amplitude $\Delta$ plays the role of the ultraviolet cut-off and temperature $T$ presents the infrared cut-off).

Some effects do not depend on the ultraviolet cut-off and are only determined by the number of the "relativistic" fermionic and bosonic fields. This allows us in particular
 to study analogs of gravitational anomalies\cite{NissinenVolovik2022,Stone2020,Laurila20} 
(including the Nieh-Yan torsional anomaly,\cite{NiehYan1982a,NiehYan1982b,Nieh2007} which acquires the universal temperature correction, $\partial_\mu j^\mu_5 =-\frac{T^2}{12} N(x)$, where $N(x)$ is the Nieh-Yan invariant);
to consider  the temperature correction to the gravitational coupling,\cite{VolovikZelnikov2003}  $K(T) -K(0) =-N_F T^2/288$, where $N_F$ is the number of fermionic species; to probe experimentally the Adler-Bell-Jackiw axial anomaly $\partial_\mu j^\mu_5 \sim {\bf B}\cdot {\bf E}$ using dynamics of vortex-skyrmions, which produces the effective magnetic ${\bf B}$ and electric ${\bf E}$ fields;\cite{Bevan1993} etc.

The general topological invariant, which supports the masslessness of fermions, is expressed in terms of the matrix Green's function in the 4-dimensional momentum-frequency space $p_\mu$:\cite{Volovik2003}
 \begin{equation}
N_3({\cal K}) = \frac{e_{\alpha\beta\mu\nu}}{24\pi^2}~
{\bf tr}\left[{\cal K}\int_\sigma   dS^\alpha
~ G\partial_{p_\beta} G^{-1}
G\partial_{p_\mu} G^{-1} G\partial_{p_\nu}  G^{-1}\right]
\label{Ksymmetry}
\end{equation}
Here the integral is over the 3D surface $\sigma$ around the singular point   in the 4-momentum space, and ${\cal K}$ is the matrix, which determines the proper discrete symmetry of the vacuum. 
For the Standard Model fermions the  discrete symmetry ${\cal K}$ is in Ref.\cite{Volovik2010}. This symmetry does not allow annihilation of the Weyl points with different chiralities. When this symmetry is broken or violated, the pair of Weyl fermions forms the massive Dirac particle. 

In this scenario, the emergent space is 3-dimensional, and the spacetime is correspondingly the 4-dimensional, which follows from the topology of Weyl point. This is one of the condensed matter answers to the questions \#25 and \#29. The emergence of the gauge fields in this scenario is the possible answer to the question \#31.

Note that in this scenario the gravitational tetrads $e_a^\mu$ emerge in the form of the contravariant vectors, as distinct from the covariant tetrads $e^a_\mu$ emerging in the Akama-Diakonov scenario in Eq.(\ref{TetradsFermions}). The same difference concerns the metric field, with $g^{\mu\nu}$ emerging in the Weyl point scenario, and $g_{\mu\nu}$ emerging in the Akama-Diakonov scenario.
This demonstrates that  in the Weyl point scenario the geometry emerges in the momentum-frequency space, while in the Akama-Diakonov scenario the geometry emerges in the real spacetime.

 \subsection{Elasticity tetrads}
 \label{gravity3}

 In these scenario, the gravitational tetrads are represented by the elasticity tetrads, which describe the elasticity theory in crystals.\cite{DzyalVol1980,NissinenVolovik2019,NissinenVolovik2018,Nissinen2022,Nissinen2023} 
 In this approach, an arbitrary deformed crystal structure can be described as a system of three crystallographic surfaces, Bragg planes, of constant phase $X^a(x)=2\pi n^a$, $n^a \in \mathbb{Z}$ with $a=1,2,3$. The intersection of the surfaces
\begin{equation}
X^1({\bf r},t)=2\pi n^1 \,\,, \,\,  X^2({\bf r},t)=2\pi n^2 \,\,, \,\, X^3({\bf r},t)=2\pi n^3 \,,
\label{points}
\end{equation}
represent the lattice points of a deformed crystal. In the continuum limit, the elasticity triads are  gradients of the phase functions:
\begin{equation}
E^{~a}_i(x)= \partial_i X^a(x)\quad i=x,y,z, \quad a=1,2,3 \,.
\label{reciprocal}
\end{equation}
The extension to the 3+1 tetrads produces the scenario, in which the quantum vacuum is the plastic (malleable) fermionic crystalline medium.\cite{KlinkhamerVolovik2019}  This vacuum crystal does not have the equilibrium value of lattice size, and thus all the deformations are possible. The curvature and torsion are produced by the topological defects of quantum crystals --  disclinations and dislocations correspondingly.\cite{Bilby1956,Kroner1960} Similar approach to gravity with four scalar fields 
$X^a(x)$ see in Ref.\cite{KoivistoZlosnik2023}.

 \subsection{Rectangular tetrads}
 \label{gravity4}
 
 The condensed matter demonstrates that metric may emerge from the non-quadratic vielbein. The dimension of spin space in the vielbein matrix can be smaller or larger than the dimension of coordinate space. Example of the latter is the $4\times 5$ vielbein with dimension 5 of the spin space, which takes place in the planar phase of superfluid $^3$He.\cite{Volovik2022} Such scenario can be  extended to the Akama-Diakonov gravity. In this scenario one may have continuous change of the signature of the metric.\cite{Volovik2022} Dynamical signature has been also discussed in Ref.\cite{Zubkov2022}. 

The condensed matter demonstrates also the opposite scenario, when the dimension of spin space in the elasticity vielbein is smaller than the dimension of coordinate space.\cite{Volovik2022}  Examples are provided by the $3\times 2$ vielbein for lattices of linear objects (vortex lattice in superfluids and superconductors and skyrmion lattice in chiral magnets) and by the $3\times 1$ vielbein for systems of planes (the smectic liquid crystal).

 \subsection{Multiple tetrads}
 \label{gravity5}

 The broken symmetry may lead to formation of different tetrads for different fermionic species.\cite{Parhizkar2022} For example, if Minkowski vacuum is degenerate with respect to discrete symmetries,\cite{Vergeles2021} one may have separate tetrads: i) tetrads for left fermions; ii) tetrads for right fermions; iii) tetrads for left antiparticles; and iv) tetrads for right antiparticles.\cite{Volovik2022a} This serves as the multi-tetrad extension of bi-metric gravity introduced by Rosen.\cite{Rosen1940}
 
 \subsection{Complex tetrads}
 \label{gravity6}

The scenario with complex tetrads takes place in the B-phase of superfluid $^3$He.\cite{Volovik1990} 
The possible realization in gravity is in Ref.\cite{Bondarenko2022}. In $^3$He-B, the collective modes of  the complex order parameter (complex tetrad) contain 14 heavy Higgs modes and one pseudo-Goldstone mode with small mass (gap).\cite{Zavjalov2016}  This suggests that in addition to the 125 GeV Higgs boson, the heavier Higgs bosons are waiting to be discovered in particle physics.\cite{VolovikZubkov2014,VolovikZubkov2015} 

 \section{Discussion of emergent gravity}
 \label{gravity7}
 
 \subsection{Tetrads are more fundamental than metric}
 
In all six scenarios of emergent gravity, tetrads are the primary emergent objects, while metric is the secondary object, which is the bilinear form of tetrads.  This would mean that geometry is the secondary emergent phenomenon. Whatever scenario (if any) is preferred by Nature, the gravity is not described by Einstein metric theory and requires the extended theory in terms of tetrads such as the Einstein–Cartan–Sciama–Kibble theory.\cite{Hehl2023} This is the condensed matter answer to the question  \#13.

There are the condensed matter scenarios, in which metric emerges as primary object, such as acoustic metric in moving liquid,\cite{Unruh1981,Visser1998} see also recent review on analogue simulations of quantum gravity with fluids.\cite{Braunstein2023}  These analogs of gravity do not describe the interaction of gravity with fermions, and cannot serve as the guiding rule for constructing the quantum gravity. But the flow acoustic metric demonstrated the advantage of the Painlev\'e-Gullstrand coordinates in general relativity.\cite{Painleve,Gullstrand} 

 \subsection{Painlev\'e-Gullstrand coordinates and their extensions}
 \label{PG}

 Painlev\'e-Gullstrand coordinates describe the metric in the whole range of radial coordinates, $0<r < \infty$, without singularities at the horizons. This allows for the analytic consideration both inside and outside the horizons.\cite{Visser2022} 
In their extended versions these coordinates have the Arnowitt-Deser-Misner form:\cite{ADM2008} 
\begin{equation}
ds^2 =- N^2dt^2 + \gamma_{ik}(dx^i - v^i dt)( dx^k -v^k dt)\,,
\label{ADMacoustic}
\end{equation}
where the shift function $v^i$ is the analog of the flow velocity in acoustic gravity (velocity of the superfluid component).
These coordinates resolve between the black and white holes by the direction of the "flow of the vacuum". One has ${\bf v}=-\hat{\bf r} \sqrt{2MG/r}$ for the flow towards the singularity in case of the black hole, and  ${\bf v}=\hat{\bf r} \sqrt{2MG/r}$ for the flow away from the singularity in case of the white hole (here $M$ is the mass of black and white holes).  This, in particular, allows to calculate the macroscopic quantum tunneling between the black and white   holes.\cite{Volovik2022e}  

In Weyl semimetals, where geometry comes from the momentum space, the shift function $v^i$ and the corresponding black hole and white hole horizons are obtained by the proper tilting of the Weyl cone in the fermionic spectrum.\cite{Volovik2016,Wilczek2020} 

The extended Painlev\'e-Gullstrand coordinates are useful also for consideration of the systems with several horizons. For the electrically charged Reissner-Nordstr\"om black hole with two horizons, where the flow (the shift function) everywhere has the same direction (towards the singularity), the extended coordinates can be found in Ref.\cite{Volovik2003b}. 

In case of the Schwarzschild black hole with mass $M$ in the de Sitter spacetime the shift function is:
\cite{Volovik2023f} 
\begin{equation}
  v(r)= \sqrt{C(1-C)} \,\sqrt{\frac{r+2r_0}{3rr_0^2}}\,\,(r-r_0)\,\,,\,\, C=3(GMH)^{2/3}\,.
\label{vSdSADM3}
\end{equation}
Here the flow changes the direction at the zero-gravity surface at $r=r_0$.
The observers at $r<r_0$ have velocity $v(r)<0$ and are free falling towards the black hole horizon and then to the central singularity, while the observers at $r>r_0$ have $v(r)>0$ and are free falling towards the cosmological horizon.

 The ionization temperature measured by the static observer at $r=r_0$ is $T=\sqrt{3}H/\pi$.\cite{Volovik2023k} It does not depend on the position of the black hole horizon. This ionization temperature is twice larger than the Hawking radiation temperature obtained by Bousso and Hawking\cite{BoussoHawking1996}  in the limit $C\rightarrow 1$, when the cosmological and black hole horizons are close to each other. The origin of this factor 2 is similar to that in the de Sitter spacetime in Sec.\ref{deSitterPerturbed}, i.e. the incomplete information received by a subluminal observer.
 
It is important that in condensed matter, the black hole and white hole are different objects, which have different directions of flow. While in general relativity they can be transformed to each other by the coordinate transformation. Such coordinate transformation has singularity. This demonstrates that for some singular coordinate transformations the general covariance does not work. Moreover, the singularity in the transformation between the black and white holes allows to find the probability of the macroscopic quantum tunneling between these objects.\cite{Volovik2022e}

 \subsection{Dimensionless physics}
 \label{dimensionless}
 
 The important property of the elasticity tetrads in Sec. \ref{gravity3} is that being the derivatives of the dimensionless phase variables, they have the  dimension of  inverse length,  $[E^{~a}_\mu]=1/[L]$. In the scenario in Sec. \ref{gravity1}, where tetrads  emerge as bilinear combination of the fermion operators, they also have the dimension $1/[L]$.\cite{VladimirovDiakonov2012}
The metric $g_{\mu\nu}$, which originates from the tetrads,
\begin{equation}
g_{\mu\nu}=\eta_{ab}E^{~a}_\mu E^{~b}_\nu\,,
\label{Metric}
\end{equation}
has dimension $[g_{\mu\nu}]=1/[L]^2$.

That is why the interval
\begin{equation}
ds^2=-g_{\mu\nu}dx^\mu dx^\nu\,,
\label{Interval}
\end{equation}
which is diffeomorphism invariant, is dimensionless. This is an example of the general rule, which works in these two scenarios of quantum gravity: all diffeomorphism invariant physical quantities are dimensionless.\cite{Volovik2021c,Volovik2022c}
The scalar curvature ${\cal R}$ and gravitational coupling $K$ in Eq.(\ref{GR}), the cosmological constant $\Lambda$, square of the Weyl tensor $C_{\alpha\beta\mu\nu}C^{\alpha\beta\mu\nu}$, particle masses $M$, etc.,  are all dimensionless.

This may suggest that the dimensionlessness can be the natural consequence of the diffeomorphism invariance, and thus can be the general property of any gravity, which emerges in the quantum vacuum. 
According to Vladimirov and Diakonov,\cite{VladimirovDiakonov2012}
the unconventional dimensions of the tetrads and fields "are natural and adequate for a microscopic theory of quantum gravity".

The nontrivial dimension of the metric suggests that metric is not the quantity, which describes the space-time, but the quantity, which determines the dynamics of effective low energy fields in the background of the microscopic quantum vacuum.
 
 \subsection{Gravity and quantum mechanics}
 
In all the condensed matter scenarios, gravity emerges from the quantum vacuum degrees of freedom, which are described by the quantum mechanics and quantum field theory.
That is why, the effective gravity is the consequence of quantum mechanics, and thus from the condensed matter point of view there is no contradiction between emergent gravity and quantum mechanics. This is the condensed matter answer to the question  \#3.  

On the other hand, being the product of quantum mechanics,  condensed matter cannot say anything definite on the possible origin of quantum mechanics and quantum field theory (the question  \#32). 
However, there is some connection between the process of spontaneous symmetry breaking in condensed matter and the measurement process in quantum mechanics.\cite{Grady1994}  Both processes are phenomena, which take place in the limit of infinite volume $V$ of the whole system (the thermodynamic limit). In finite systems or in the finite volume the quantum mechanics is reversible, and also in the finite system one cannot resolve between different phases of condensed matter systems, i.e. there is no second order phase transition. In condensed matter, the disjoint Gibbs distributions are reached only in the thermodynamic limit, $V\rightarrow \infty$.\cite{Sinai1983} In quantum mechanics, this limit leads to the disjoint distributions of macroscopically different quantum states with vanishing superposition of the wave functions. 

An interesting example of the spontaneous formation of the classical macroscopic state in the thermodynamic limit is the spontaneously formed coherent precession of the quantum system of spins in $^3$He-B.\cite{BunkovVolovik2013}  All spins precess with the same frequency $\omega$ and the same phase $\alpha$ without external drive (if the spin-orbit coupling is neglected).
 This corresponds to the classical rotation of the macroscopic angular momentum of the whole quantum system with the angular velocity $\Omega=\omega$ and phase (angle) $\alpha$:
 \begin{equation}
S_x+iS_y = \sqrt{S^2 -S_z^2}\, e^{i\omega t + i\alpha}\,.
\label{precession}
\end{equation}
The precessing states are degenerate with respect to  $\alpha$, but each of them represents the ground states of the quantum system at large number of magnons. This is a kind of quantum time crystal,\cite{Wilczek2012,Autti2021} which emerges in the thermodynamic limit $S_z\rightarrow \infty$.

Another possible connection relates the Planck constant $\hbar$ and the gravitational tetrads.\cite{Volovik2023a} Since $\hbar$ is not dimensionless, it cannot be the fundamental constant. 
In the dimensionless physics scenarios in Sec. \ref{dimensionless}, the Planck "constant" $\hbar$ is the element of tetrads  and has dimension of length.

 \subsection{Emergent gravity suggests extension of Standard Model}

Scenarios in Sections \ref{gravity1} and \ref{gravity2} suggest very different origins of quantum gravity. However, they have similar predictions for the number of fermionic degrees of freedom in quantum vacuum.
In the Diakonov theory, where the original action is the product of 8 fermionic operators, the grand unification with symmetry $SO(16)$ is suggested. This fits four generations of the Standard Model with 16 Weyl fermions in one generation.\cite{Diakonov2011} 
This in particular suggests that dark matter particles can be neutrino of the fourth generation,\cite{Volovik2003a} which corresponds to the scenario known as asymmetric dark matter.\cite{Petraki2013} 

The Weyl point scenario is realized, in particular, in superconductors of class 
$O(D_2)$.\cite{VolovikGorkov1985,Volovik2017}  In this superconductor, there are 8 Weyl points in the energy spectrum, which form cube in 3D momentum space giving rise to 8 Weyl fermions. In the 4D extension\cite{Creutz2008,Creutz2014} the Weyl nodes may form the 4D cube in the momentum-frequency space, which results in 16 Weyl fermions. All fermions become massive at low energy, when the symmetry ${\cal K}$ in Eq. (\ref{Ksymmetry}), which supports masslessness, is broken. That is why all neutrinos are massive Dirac particles, which is the condensed matter answer to question  \#10.

While the Weyl point scenario does not support the supersymmetry, the smallness of masses of observed particles follows from the exponential suppression of the temperature of the symmetry breaking phase transitions, which is typical in condensed matter systems. This is the condensed matter answer to question  \#11.

The numbers $2^N$ also follow from the topological analysis of condensed matter systems.
If the vacuum of Standard Model is considered as topological Weyl semimetal, the maximal number of massless fermions in the symmetric phase is $16g$, where $g$ is number of generations.
\cite{VolovikZubkov2017}  The group $Z_{16}$ also appears in the classification of topological phases in 3+1 dimension.\cite{Schnyder2016} 
All this can be in favour of the Pati-Salam type extension of the Standard Model. In particular, the Standard Model with its $G(2,1,3)$ group can be extended to grand unification $G(2,2,4,4)$ group, which may include the left and right $SU(2)$ gauge fields, the $SU(4)$ color fields, and probably the $SU(4)$ family fields. 

 \subsection{Topological objects in condensed matter and in cosmology}
 
Symmetry breaking scenario in Sec. \ref{gravity1} suggests new topological objects, such as cosmic domain walls, in which either the space components of tetrads change sign, or time component
changes sign, or the whole tetrad changes sign.\cite{Volovik2020c} In $^3$He-B such walls are the
analogs of the Kibble-Lazarides-Shafi cosmic walls bounded by strings.\cite{Kibble1982,Makinen2019}
Such combined object demonstrates two different routes to the Alice looking-glass world (anti-spacetime): the safe route around the Alice string (half-quantum vortex) and the very dangerous route through the Kibble-Lazarides-Shafi cosmic wall.\cite{EltsovNissinen2019} This concerns the question  \# 27.

 The other analogs of composite cosmological objects, such as Nambu strings bounded by monopoles -- "magnetic quarks"\cite{Lazarides2021} -- appear in $^3$He-A.\cite{Volovik2020b} 
The broad connections between the observed topological objects in condensed matter and the predicted cosmological objects is the answer to question  \# 18. 

Also, the chemistry of formation of topological "magnetic quarks" connected by "cosmic strings"  in $^3$He-A is similar to chemistry  of the formation of the multi-quark objects. Examples are the pentaquarks, which are formed by valence quarks and antiquarks joined by the strings.\cite{Andreev2023,Andreev2023b} This concerns question \# 15.

\section{Conclusion}
\label{Conclusion}

Here we considered several answers to several fundamental questions posted in Ref.\cite{Perspective}.
These answers are based on the condensed matter experience, where we know both the many-body phenomena emerging on the macroscopic level and the microscopic (atomic) physics, which generates this emergence. It is not surprizing that the same macroscopic phenomenon may be generated by essentially different microscopic backgrounds. This points to various possible directions of studying the deep quantum vacuum. All this represents the connection between the formalism of physics and the reality of the condensed matter experience (question \# 34).

\end{document}